\documentclass[pre,twocolumn]{revtex4}

\usepackage{amsmath}
\usepackage{epsfig}
\usepackage{graphicx}

\begin{document}

% Macros
\newcommand{\ex}[1]{\mathrm{e}^{#1}}
\newcommand{\MK}{\mathbf{M}} % Notation for a Markov Matrix
\newcommand{\m}{m} % Notation for the elements of the Markov Matrix
\newcommand{\from}{\leftarrow}
\renewcommand{\d}{\mathrm{d}}
\newcommand{\e}{\mathrm{e}}
\newcommand{\set}[1]{\lbrace#1\rbrace}
\newcommand{\av}[1]{\overline{#1}}
\newcommand{\eref}[1]{(\ref{#1})}
\newcommand{\etal}{{\it{}et~al.}}
\newcommand{\defn}{\textit}
\newcommand{\cF}{\mathcal{F}}
\newcommand{\cS}{\mathcal{S}}
\newcommand{\BK}{\mathbf{B}}
\newcommand{\ee}{\mathbf{e}}
\newcommand{\II}{\mathbf{I}}
\newcommand{\PP}{\mathbf{P}}
\newcommand{\DD}{\mathbf{D}}
\newcommand{\lam}{\lambda}
\newcommand{\xx}{\mathbf{x}}
\newcommand{\yy}{\mathbf{y}}
\renewcommand{\r}{\vec{r}}
\newcommand{\M}{\mathbf{P}}
\newcommand{\XX}{\mathbf{X}}
\newcommand{\YY}{\mathbf{Y}}
\newcommand{\LL}{\mathbf{L}}
\newcommand{\RR}{\mathbf{R}}
\newcommand{\pp}{\mathbf{p}}
\newcommand{\dgr}{^{\circ}}
\newcommand{\km}{\mathrm{km}}

% Style parameters
\newlength{\figurewidth}
\setlength{\figurewidth}{0.50\columnwidth}
\setlength{\parskip}{0pt}
\setlength{\tabcolsep}{6pt}
\setlength{\arraycolsep}{2pt}

\title{The Eigenmode Analysis of Human Motion}
\author{Juyong Park}
\affiliation{Department of Physics, Kyung Hee University, Seoul 130-701, Republic of Korea}
\affiliation{School of Physics and Astronomy, Seoul National University, Seoul 151-747, Republic of Korea}
\author{Deok-Sun Lee}
\affiliation{Departments of Natural Medical Sciences and Physics, Inha University, Incheon 402-751, Republic of Korea}
\author{Marta C. Gonz\'alez}
\affiliation{Department of Civil and Environmental Engineering, Massachusetts Institute of Technology, Cambridge, MA 02139, USA}
\begin{abstract}
{
Rapid advances in modern communication technology are enabling the accumulation of large-scale, high-resolution observational data of spatiotemporal movements of humans. Classification and prediction of human mobility based on the analysis of such data carry great potential in applications such as urban planning as well as being of theoretical interest.  A robust theoretical framework is therefore required to study and properly understand human motion. Here we perform the eigenmode analysis of human motion data gathered from mobile communication records, which allows us to explore the scaling properties and characteristics of human motion.}
\end{abstract}
\maketitle

\section{Introduction}
Thanks to rapid advances in data retrieval technology coupled with the advent of the Internet and personal electronic devices, we are witnessing the emergence of unprecedented opportunities to collect, analyze, and understand massive data encoding a variety of human behaviors, prompting calls for collaborative efforts from various disciplines in the scientific community~\cite{Nature:2009}. One topic gaining heightened interest in the physics community in particular is how one can understanding the spatiotemporal movements of humans~\cite{Brockmann:2006}, an example being a recent numerical study by Gonz\'alez~\etal~of human mobility data gathered from mobile phone communication records~\cite{Gonzalez:2008}.  One intriguing finding in their work was that the distance or area covered by human trajectory appears to be bounded or increase extremely slowly in time $t$, which highlights the stark difference between classical models of particle motion such as Random Walk and L\'evy Flight~\cite{Mantegna:1994,Klafter:1996} where the expected fluctuation in the displacement of a particle grow unbounded as polynomials of time $t$ (a notable exception is the Random Walk in Random Enviroment (RWRE) where a weak spatial randomness assigned to lattice points lead to a $\sim\log^2t$ behavior in the average fluctuation~\cite{Hughes:1995,Hughes:1996,Sinai:1983,Durrett:1986,Kesten:1986}).  Given the success of these models in explaining real physical systems such as diffusive gases in free space, it is natural to examine their validity in understanding human mobility data. But as Gonz\'alez~\etal~ showed, the simplistic nature of these models are not suited for that task, which showcases the inherent complexity of human motion. For the same reason, devising a simple generative model of human motion appears extremely difficult if at all possible, requiring more theoretical approaches to understanding the observed behavior of human motion on a deeper level.

As a first step in that direction, in this paper we study analytically a number of observations given in Gonz\'alez~\etal~Specifically, we employ the framework of Markov processes, the fundamental and indispensable methodology in the study of stochastic processes in a variety of scientific and engineering fields such as including linguistics, computational biology, graph theory, and information theory, to name a few\cite{Markov:2006,Shannon:1948,Bremaud:1991,Krogh:1994}: we construct the Markov transition matrix from mobile communication data, and investigate its properties via eigenmode analysis. We show that this allows us to understand the aforementioned slow increase of fluctuations in human motion in time and, furthermore, and promises to be a methodology to characterize patterns of individuals' motions according via the eigenmodes of the transition matrices.

This paper is organized as follows. In Sec.~\ref{sectionA}, we introduce the Markov transition matrix constructed from the observed transition rates between spatial coordinates. In Sec.~\ref{sectionB}, we investigate the various properties of eigenmodes, from which we present a theoretical understanding of the radius of gyration. In Sec.~\ref{sectionC}, we show that the eigenmodes act as a concise descriptor of the characteristics of individual movement patterns. Finally, we summarize and discuss our findings and their implications in Sec.~\ref{sectionD}.

\section{The Human Mobility Data and the Markov Chain Framework}
\label{sectionA}
\begin{figure}
\resizebox{1.8\figurewidth}{!}{\includegraphics{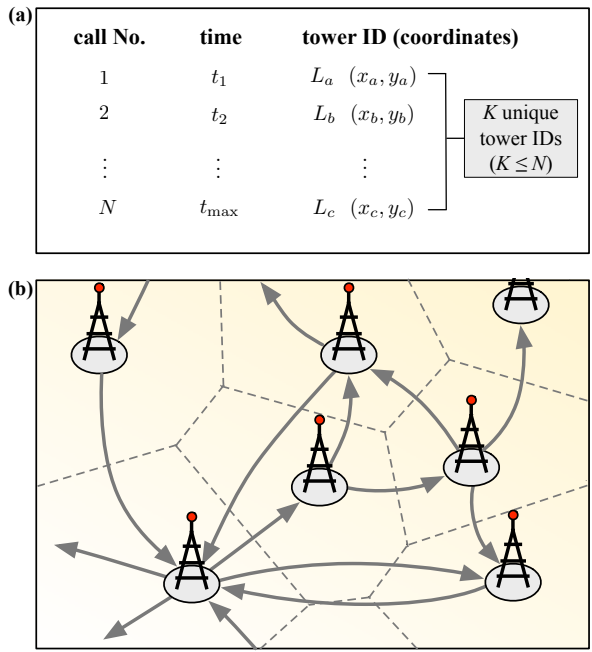}}
\caption{(a) The format of data used in our analysis. For each mobile user, a table lists the time and location, represented by the nearest transmission tower IDs, of $N$ calls initiated by the individual. The number of unique tower IDs, $K$, satisfies $K\le N$. (b) A graphical representation of the data. Transmission towers distributed over the 2-D space constitute a mesh of non-overlapping \emph{Voronoi cells}.}
\label{data_scheme}
\end{figure}
For this study, we employed a mobile phone-usage data set provided from a western European service provider, used in a previous study of Gonz\'alez~\etal~\cite{Gonzalez:2008}. Fig.~\ref{data_scheme}~(a) illustrates the basic format of the data; for each individual we have a table listing the time and the location of each of $N$ calls initiated by the user (varying from individual to individual) during an observational period of six months. The spatial location of the user at the call initiation is given by the ID of the transmission tower nearest to the user, meaning that the resolution of the user location is dictated by the size of non-overlapping \emph{Voronoi cells} centered around each tower, presented schematically in Fig.~\ref{data_scheme}~(b). Since one Voronoi cell corresponds to one transmission tower, we may use the two concepts interchangeably. The number $K$ of unique tower IDs appearing in Fig.~\ref{data_scheme}~(a) satisfies $K\le N$, the number of calls.  From this table, we build for each individual a $K\times K$ \textbf{Markov matrix }(also called the transition matrix) $\MK=\{m_{ij}\equiv m_{i\from j}|1\le i,j\le K\}$, where $m_{ij}$ is the probability that the user has been observed (made a call) at Voronoi cell $i$ right after being observed at cell $j$ with no other observation in between: if the individual was observed at $j$ for a total of $n_j$ times and subsequently at tower $i$ with no observations in between for a total of $n_{ij}$ times ($0\le n_{ij}\le n_j$) during the observation period, $\m_{ij}\equiv n_{ij}/n_j$~\footnote{In order to avoid the last cell visited at $t_{\mathrm{max}}$ becoming a ``sink'' which happens when there is no record of transition away from it, we assume that the individual makes an additional transition from it to the first Voronoi cell visited at $t_1$}.

The matrix $\MK$ constructed in this manner exhibits the following mathematical properties.  First, each column of $\MK$ sums to unity since $\sum_i\m_{ij}=\sum_in_{ij}/n_j=1$. The same cannot be said for the rows, however, since $\MK$ is asymmetric in general.  Second, from the Perron-Frobenius theorem, $\MK$ always has a leading eigenvalue $\lambda_0\equiv1$, whose associated right eigenvector is $\ee_0=\pp$, where $\pp=\set{p_1,p_2,\ldots,p_K}$ represents the stationary probability of observing the user at each tower location ($p_i=n_i/N$). This can be easily understood: the stationarity condition is $p_i = \sum_j m_{ij}p_j$, which we can also write as $\pp=\lambda\MK\pp$ with $\lambda=1$.  Lastly, all eigenvalues satisfy $|\lambda_i|\le 1$, i.e. they  are confined within the unit circle on the complex plane.

Our presentation of human motion as a stochastic Markov process involves some simplification. Firstly, we assume that the calls were placed evenly spaced throughout the six-month observation period. Therefore, we do not address the impact of the the inter-call time distribution~\cite{Candia:2008} or the possible explicit time dependence of the element $m_{ij}$ of $\MK$.  Secondly, our $\MK$ incorporates the history dependence of one step only: a transition depends only on the current location, but not further into the past. Without these simplifications, one could possibly carry out a much more detailed and precise analysis. We may, for instance, construct a time-dependent matrix $\MK(t)$ reflects the changes in call patterns during the course of the day. We may also construct $\MK$ to be of a higher dimension (e.g., $K\times K\times K$) so that it incorporates history further into the past than one step as we do here.  But doing so would render the analysis presented in this paper unnecessarily complex.  Nevertheless, our current construction of $\MK$ still yields interesting and useful insights to understanding human mobility, which we hope lays a foundation for future research that incorporate those possibilities.

\section{Time-evolution of the radius of gyration}
\label{sectionB}
\begin{figure}
\resizebox{2.0\figurewidth}{!}{\includegraphics{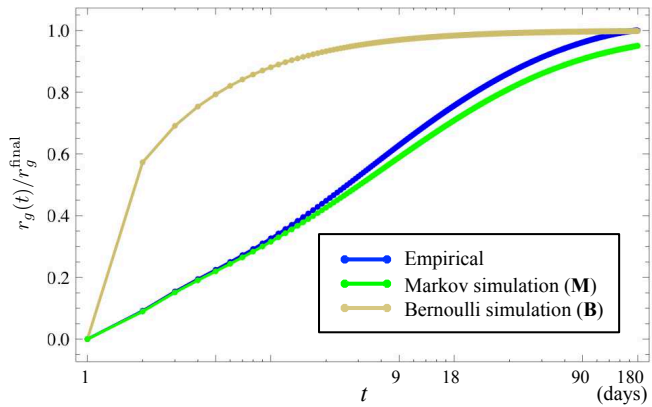}}
\caption{Comparison of the evolution of the radii of gyration (normalized) between actual data (blue) and simulations using the Markov matrix $\MK$ (green) and the Bernoulli matrix $\BK$ (yellow). $\MK$ exhibits a markedly closer reproduction of the actual data, indicating that incorporating history (in the case of $\MK$, one step) is essential in modeling human mobility.}
\label{rgaverage}
\end{figure}

A quantity of central focus in Gonz\'alez~\etal~is the \emph{radius of gyration}, defined in a squared form as~\cite{Goldstein:1950}
\begin{align}
r_g^2(s) \equiv \frac{1}{s}\sum_{l=1}^s\biggl|\vec{r}(l)-\frac{\sum_{k=1}^s\vec{r}(k)}{s}\biggr|^2~~~~(1\le s\le N)
\label{defrg}
\end{align}
a dynamic variable (as a function of $s$, the index of the calls in Fig.~\ref{data_scheme}~(a)) that measures the size of the trajectory at each transition, which demonstrated the stark contrast between actual human motion and classical random walk models. We here ask whether our stochastic process approximation is able to reproduce its slow, sub-polynomial growth shown in Ref.~\cite{Gonzalez:2008}.  In Fig.~\ref{rgaverage} we compare, the $x$-axis rescaled from $s$ to real time $t$, the empirical temporal growth of $r_g(t)/r_g^{\textrm{final}}$ and that from simulations based on $\MK$, averaged over all individuals with $r_g^{\textrm{final}}\ge100\km$.  We see that the actual curve and the simulation curve match within $\sim 5\%$, validating our approximation.  To highlight the importance of incorporating history we also plotted the corresponding curve generated from the \textbf{Bernoulli matrix} $\BK$. A Bernoulli matrix is also a transition matrix with no history dependence: its entry $b_{ij}$ is simply equal to $p_i=n_i/N$, simply the stationary probability  to be at $i$ with no regard to one's present location.  $\BK$ shows a markedly quicker convergence of $r_g$ to $r_g^{\textrm{final}}$, significantly deviating from actual data.  This handily demonstrates that temporal correlation between individual's locations is an essential factor in modelling human mobility. In the following, we study the relationship between $r_g(s)$, Eq.~\eref{defrg} and the transition matrix to better understand the origin of this behavior.

\subsection{Radius of Gyration and the Transition Matrix}
Under the Markov process $\MK$, the expected value of the square of the radius of gyration, Eq.~\eref{defrg}~\footnote{We study the squared form $r_g^2$ rather than $r_g$ which is much more complicated to treat analytically.}, at the $s$-th step can be written as
\begin{align}
	\av{r_g^2(s)} &= \frac{1}{s}\sum_{l=1}^s \av{\bigl|\r(l)\bigr|^2}-\frac{1}{s^2}\sum_{l,k}^s\av{\r(l)\cdot\r(k)} \nonumber \\
		&= \biggl(\frac{1}{s}-\frac{1}{s^2}\biggr)(\xx^T\PP\xx+\yy^T\PP\yy) \nonumber \\
		&{}{}-\frac{2}{s^2}\sum_{s'=1}^s(\xx^T\MK^{s'}\PP\xx+\yy^T\MK^{s'}\PP\yy), 
\label{rgdef}
\end{align}
where $\r(l)$ denotes the location of the user at the $l-$th step ($\le s$), $\PP$ is a diagonal $K\times K$ matrix of $P_{ij}=\set{p_i}\delta_{ij}$, and $\xx=\{x_1,\ldots,x_K\}$ and $\yy=\{y_1,\ldots,y_K\}$ are the coordinates of the $K$ unique Voronoi cells that a user has visited.  Thanks to the stationary condition that $p_i$ represents the probability of finding the individual at cell $i$ at any step, the vector product $\overline{\vec{r}(l)\cdot\vec{r}(l+s')}$ depends only the step difference $s'$, leading to $\av{\vec{r}(l)\cdot\vec{r}(l+s')}=\av{\vec{r}(1)\cdot\vec{r}(1+s')}=\xx^T\MK^{s'}\PP\xx+\yy^T\MK^{s'}\PP\yy$.

The transition matrix can be expressed as $\MK=\LL^{-T}\DD\RR$, where $\LL$ and $\RR$ are its left- and right-eigenvector matrices normalized so that $\RR\LL^T=\II$, and $\DD$ is the diagonal matrix of $\MK$'s eigenvalues $\set{\lambda_k|1\le k\le K}$. Therefore we can rewrite Eq.~\eref{rgdef} as
\begin{align}
\av{r_g^2(s)} &= \sum_{k=1}^K(a_x^kb_x^k+a_y^kb_y^k)\biggl[\bigl(1-\frac{1}{s}\bigr)-\frac{2}{s^2}\sum_{{s'}=1}^s(s-{s'})\lam_k^{s'}\biggr] \nonumber \\
&= \sum_{k=1}^K\rho_k\biggl[1-\biggl(\frac{1+\lam_k}{1-\lam_k}\biggr)\frac{1}{s}+\biggl(\frac{2\lam_k(1-\lam_k^s)}{(1-\lam_k)^2}\biggr)\frac{1}{s^2}\biggr]\nonumber \\
&\equiv \sum_k\rho_k\cF(\lam_k,s),
\label{decomposition}
\end{align}
where we have defined the \emph{mode weight} $\rho_k\equiv(a_x^kb_x^k+a_y^kb_y^k)$, invariant in $s$, with
\begin{eqnarray}
\RR^{-1}\PP\xx = \set{a^x_k}&~~~~\mathrm{and}~~~~& \xx^T\LL^{-T} = \set{b^x_{k}}^T\nonumber \\
\RR^{-1}\PP\yy = \set{a^y_k}&~~~~\mathrm{and}~~~~& \yy^T\LL^{-T} = \set{b^y_{k}}^T.
\end{eqnarray}

\begin{figure}
\resizebox{2.0\figurewidth}{!}{\includegraphics{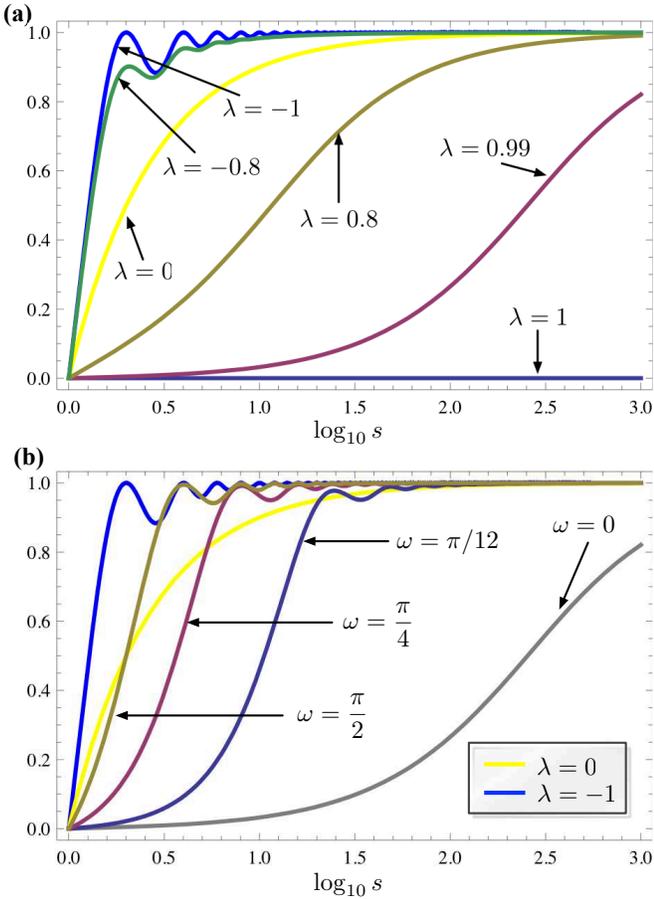}}
\caption{The increasing behavior of the real part of $\cF(\lam,s)$ for various $\lambda$ values. (a) When $\lambda$ is real, $\cF(\lambda,s)$ increases slower as $\lambda\to1$, while for $\lambda_0\equiv1$, $\cF(1,s)\equiv0$. (b) The tendency is similar in the case of complex $\lam$ (shown here for $\lambda=0.99\ex{i\omega}$), showing a faster convergence to $1$ as $\omega$ is tuned away from $0$.}
\label{Fbehavior}
\end{figure}

Thus we have decomposed the expected value of $r_g^2(s)$ into a sum of independent contributions from the $K$ eigenmodes of $\MK$.  Specifically, Eq.~\eref{decomposition} says that each eigenmode contributes in two ways: via the $s$-independent weight $\rho_k$ and the $s$-dependent $\cF(\lambda_k,s)$. We show in Fig.~\ref{Fbehavior} the behavior of $\cF(\lambda,s)$\footnote{In Fig.~\ref{Fbehavior} we show only the real portion of $\cF$: in truth, $\textrm{Im}[\cF]\ne0$ when $\textrm{Im}[\lambda]\ne0$, i.e. complex. In that case, its conjugate $\lambda^*$ is also an eigenvalue of $\MK$ with mode weight $\rho^*$. The combined effect of the two are basically $\rho\cF(\lambda,s)+\rho^*\cF(\lambda^*,s)=2Re[\rho\cF(\lambda,s)]$.}: (a) When $\lambda$ is real (i.e. $\omega\equiv\arg(\lambda)=0~\textrm{or}~\pi$), the closer $\lambda$ is to $-1$, the faster the convergence of $\cF(\lambda,s)$ to its maximum value $1$. (Interestingly, $\lambda_0=1$ does not contribute at all, since $\cF(1,s)\equiv0$. Also, the Bernoulli matrix $\BK$ has $\lambda=0$ as its only eigenvalue except the leading $\lambda_0=1$. Compare the yellow curve in Fig.~\ref{rgaverage} and the curve for $\lambda=0$.) (b) When $\lambda$ is complex (i.e. $\omega=\arg(\lambda)\ne0,~\pi$), convergence to the maximum is also accelerated as $\omega$ approaches $\pi$ for a given value of $|\lambda|$ ($|\lambda|=0.99$ in Fig.~\ref{Fbehavior}~(b)). 

\subsection{Eigenmode weight $\rho$}
Comparing Figs.~\ref{rgaverage}~and~\ref{Fbehavior}, we find that typical temporal behavior of $r_g$ is most likely due to modes with eigenvalues that are real and positive, close to $1$.  Thus there are two possible scenarios: The first is that most eigenvalues of $\MK$ happen to be $\simeq1$, resulting in a slow increase of $r_g$ \emph{regardless} of any other factors in Eq.~\eref{decomposition} such as $\xx$ and $\yy$. The other is that, while the eigenvalues are broadly distributed over the complex unit circle, only those that are close to $1$ carry large mode weights $\rho$.

\begin{figure}
\resizebox{2.0\figurewidth}{!}{\includegraphics{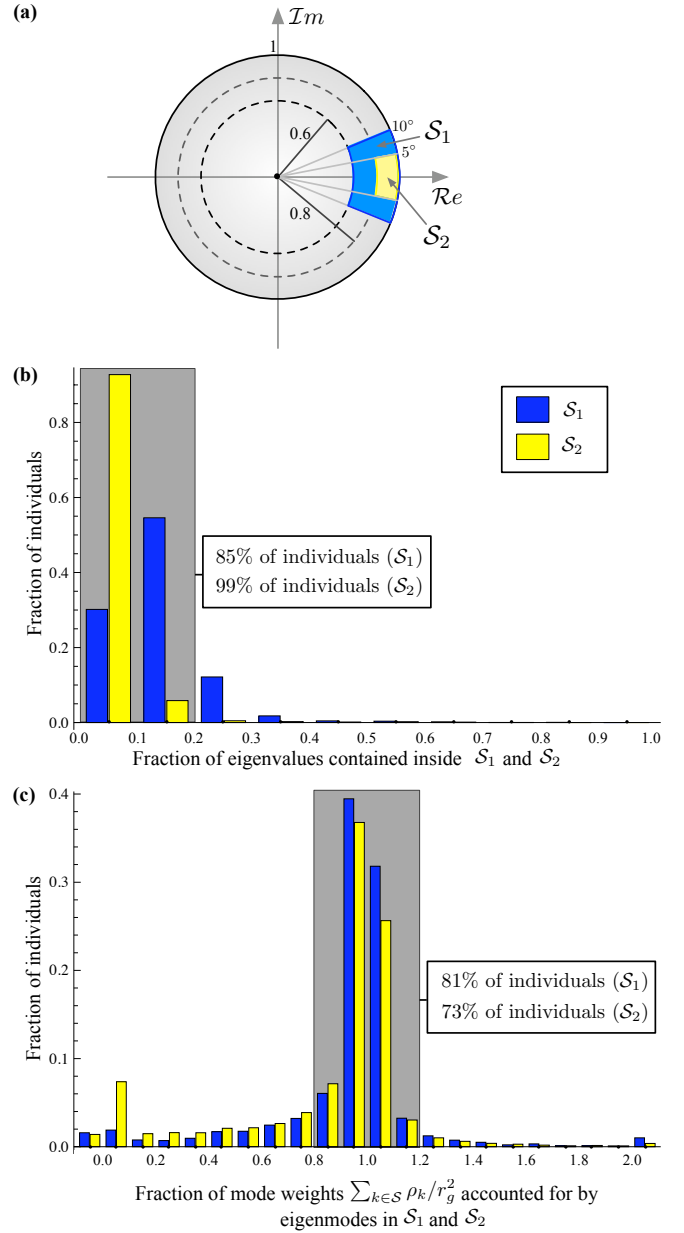}}
\caption{(a) Eigenvalues of $\MK$ are distributed over the unit circle on a complex plane. We define two sets $\cS_1$ and $\cS_2$ of eigenvalues near $1$ for each individual contained in the two colored areas inside the unit circle, excluding $\lambda_0\equiv1$. Note that $\cS_1\supset\cS_2$) (b) Histogram of the fraction of an individual's eigenvalues contained in $\cS_1$ and $\cS_2$. For $85\%$ and $99\%$ of individuals, $\cS_1$ and $\cS_2$ respectively contain fewer than $20\%$ of their eigenvalues, indicating the broad distribution of eigenvalues over the complex unit circle. (c) Histogram of the fraction of total mode weights from eigenmodes in $\cS_1$ and $\cS_2$, $\sum_{k\in\cS}\rho_k/r_g^2$. For $81\%$ and $73\%$ of individuals, eigenmodes in $\cS_1$ and $\cS_2$ account for $80\%-120\%$ of $r_g^2$, indicating their importance in the dynamics of the radius of gyration.}
\label{RkHistogram}
\end{figure}

To find the more likely scenario, for each individual we investigated the following two sets $\cS_1$ and $\cS_2$ of eigenvalues:
\begin{itemize}
	\item Set $\cS_1$ comprises eigenvalues with $|\lambda|\ge0.6$ and $-10\dgr\le\arg(\lambda)\le10\dgr$, contained within the blue and the yellow 
	areas inside the unit circle in Fig.~\ref{RkHistogram}~(a).
	\item Set $\cS_2$ comprises eigenvalues with $|\lambda|\ge0.8$ and $-5\dgr\le\arg(\lambda)\le5\dgr$, contained within the yellow area in Fig.~\ref{RkHistogram}~(a).
\end{itemize}

If the second scenario is correct, we should see only a small number of eigenmodes contained within $\cS_1$ and $\cS_2$, while their combined mode weights $\sum_{k\in\cS}\rho_k$ account for most of $\sum_k\rho_k=r_g^2$. This is precisely what we see in Fig.~\ref{RkHistogram}~(b) and (c): in Fig.~\ref{RkHistogram}~(b), we see that for 85\% and 99\% of the individuals, respectively, $\cS_1$ and $\cS_2$ contain fewer than $20\%$ of their eigenvalues. On the other hand, Fig.~\ref{RkHistogram}~(c) shows that, for $81\%$ and $74\%$ of the individuals, respectively, eigenmodes in $\cS_1$ and $\cS_2$ account for $80\%-120\%$ of $r_g^2$ (note $\rho$ need not be positive for every eigenmode -- $\rho$ only has to sum up to $r_g^2$, so the sum of $\rho$ of a subset of eigenmodes may exceed $r_g^2$).  Thus we can conclude that only a minority of eigenvalues near $1$ account for most of $r_g^2$, causing the slowly increasing behavior we see in Fig.~\ref{rgaverage}. 

Next, we further explore the implications of such behavior at the individual level by understanding the meaning of the eigenmodes of $\MK$ in detail.

\section{The Meaning of Eigenmodes}
\label{sectionC}
Let $p_k(s)$ denote the probability to find an individual at a Voronoi cell $k$ at step $s$ under the Markov process. Then we can express $\pp(s)=(p_1(s),p_2(s),\ldots,p_K(s))$ as
\begin{align}
\pp(s) = \MK^s\pp(0) = \sum_{k=1}^Ka_i\lambda_k^s\ee_k,
\label{Prob}
\end{align}
where $\set{\ee}$ are the right eigenvectors of $\MK$. The coefficients $\set{a_k}$ can be obtained from $\RR^{-1}\pp(0)=(a_1,a_2,\ldots,a_n)$. This means that $\pp(s)$ is a sum of independent temporal evolutions of the eigenmodes themselves. Again, note that $a_k$, $\lambda_k$, and $\ee_k$ can be complex due to the asymmetric nature of $\MK$. A complex-valued eigenmode would contribute to $\pp(s)$, in conjunction with its complex conjugate (which is also an  eigenmode of $\MK$), as
\begin{align}
	a_k\lambda_k^{s}&\ee_k + a_k^*(\lambda_k^*)^{s}\ee_k^*\nonumber \\
	&\propto |a_k||\lambda_k|^{s}\bigl(\ldots,~|e_{k,n}|\cos(\alpha_k+\omega_k s+\phi_{k,n}),~\ldots\bigr),
\label{pevolve}
\end{align}
where $e_{k,n}$ is the $n$-th component of $\ee_k$, and $\alpha_k\equiv\arg(a_k)$, $\omega_k\equiv\arg(\lambda_k)$, and $\phi_{k,n}\equiv\arg(e_{k,n})$, guaranteeing that $\pp(s)$, a probability, is always real.

In the asymptotic $s\to\infty$ limit all modes decay to zero except the leading mode with $\lambda_0\equiv1$, so that $\pp(\infty)=\pp=\set{p_i}$, the stationary occupation probability.  For finite $s$, however, the transient dynamics, governed by non-leading eigenvalues, can be important and offer interesting detail of the physical process. In this section, therefore, we study the non-leading eigenmodes, and discuss how they can characterize individual human mobility patterns. For convenience, we discuss real- and complex-valued eigenmodes separately.

\subsection{Real-Valued Eigenmodes}
\begin{figure*}
\resizebox{4.0\figurewidth}{!}{\includegraphics{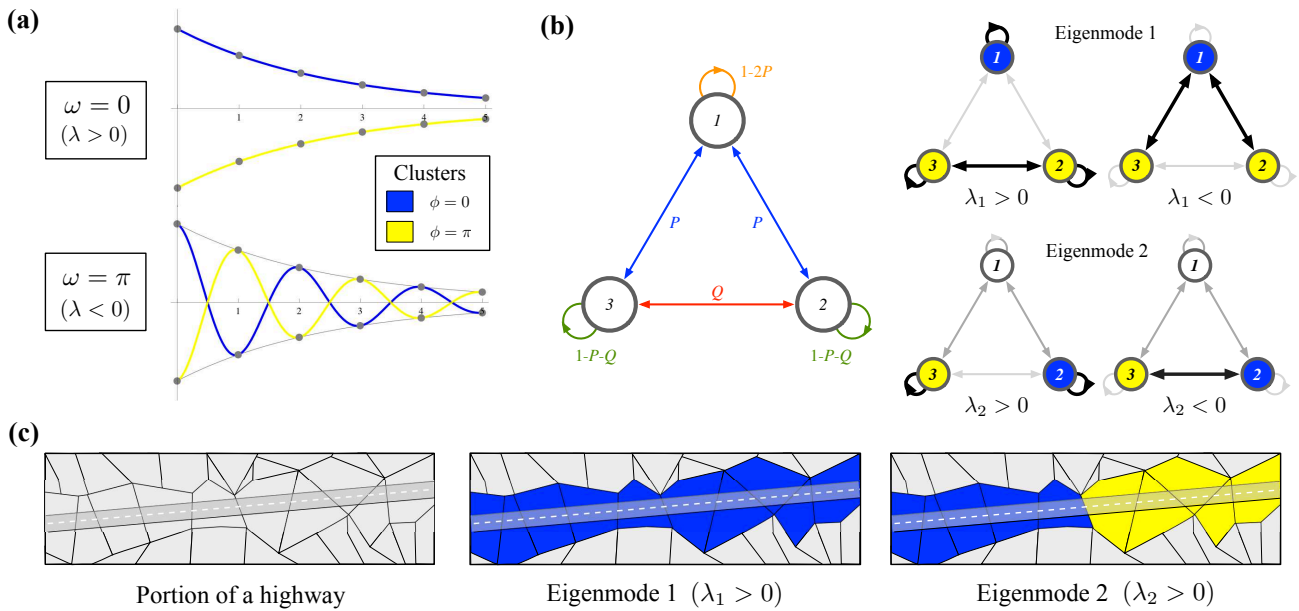}}
\caption{(a) The two cases of the temporal evolution of the occupation probabilities in eigenmodes with real $\lambda$. When $\lambda>0$ ($\omega=0$), the occupation probability of a Voronoi cell attenuates monotonically towards the stationary state, meaning a small transition between the different clusters (blue and yellow). When $\lambda<0$ ($\omega=\pi$), the occupation probabilities fluctuate sinusoidally, indicating an active exchange of probability flux between the two clusters (the gray envelope curve is a guide). (b) A simple Markov process consisting of three cells with symmmetric transition probabilities, along with a diagrammatic summary of the dynamic properties of its eigenmodes given in Eq.~\eref{mc1sol}. In eigenmode 1, cell 1 constitutes one cluster (blue) and cells 2 and 3 the other cluster (yellow). In accordance with (a), when $\lambda_1>0$, the individual's motion is mostly confined to inside one cluster, whereas when $\lambda_1<0$, it transitions often between the two clusters. In eigenmode 2, cells 2 and 3 constitute two distinct clusters (cell 1 is irrelevant, with its eigenvector component $0$ in $\ee_2$). Thus eigenmode 2 presents a finer detail of motion between cells 2 and 3 which is absent in eigenmode 1. (c) A real-world example of Voronoi cell clustering. A portion of a major highway is contained within ten Voronoi cells (left).  According to the eigenmode 1 ($\lambda_1>0$) of an individual who traveled along these cells, the entire portion of the highway belongs to a single cluster. However, this does net tell us whether they were traversed in a linear fashion or in a haphazard way (however unlikely). According to the eigenmode 2, however, the highway is divided into two adjacent compact clusters. In accordance with (b), $\lambda_2>0$ tells us that there were minimal transitions between the clusters, indicating that the cells were traversed in a linear fashion, which agrees with our intuition.}
\label{realeig}
\end{figure*}

Note that Eq.~\eref{pevolve} holds for a real-valued eigenmode as well: the angular variables $\set{\alpha_k,\omega_k,\phi_{k,n}}$ are merely restricted to either $0$ or $\pi$. First, consider $\phi_{k,n}$. Within the same eigenmode $k$, it is possible that some $\phi_{k,n}$ are $0$ while others are $\pi$ (equivalently, the components can be positive or negative). Since each component corresponds to a Voronoi cell, this means that we can cluster the Voronoi cells into two distinct groups, according to the sign of their eigenvectors (only in regards to this specific eigenmode; other eigenmodes can offer different groupings). According to Eq.~\eref{pevolve}, this means that as $s$ is incremented, the occupation probabilities of Voronoi cells in the same cluster are \emph{in phase}, and out of phase with cells in the other cluster. Additionally, if $\lambda_k>0$ ($\omega_k=0$), the occupation probabilities attenuate monotonically, while if $\lambda_k<0$ ($\omega_k=\pi$), they fluctuate as they attenuate. This behavior is visualized in Fig.~\ref{realeig}~(a), showing the occupation probabilities of the two said clusters (blue and yellow)~\footnote{$\alpha_k$, dependent only on the initial condition, merely dictates the relative magnitude of each eigenmode in Eq~\eref{pevolve} at $s=0$.}

We can rephrase this in a more intuitive way: When $\lambda_k>0$ ($\omega_k=0$), the individual is likely to be contained within the same cluster as time progresses, with a small probability of transitioning into the other cluster.  If the individual does cross over to the other cluster, however, the trend persists: their motion is now likely to persist in that group. On the other hand, when $\lambda_k<0$ $(\omega_k=\pi)$, at every step the individual tends to transition to a cell of the other cluster, resulting in a large probability flux exchange between the clusters (hence the fluctuation). 

A simple illuminating example is given in Fig.~\ref{realeig}~(b). In it we have three Voronoi cells represented as nodes $1$, $2$, and $3$. The corresponding transition matrix $\MK$ is
\begin{align}
\MK=
\left(
	\begin{array}{ccc}
	1-2P	& P	& P \\
	P	& 1-P-Q	& Q \\
	P	& Q	& 1-P-Q
	\end{array}
\right),
\end{align}
which results in the following three eigenmodes:
\begin{align}
	\textrm{Eigenmode 0} :&~\lambda_0=1,~\ee_0=\frac{1}{3}(1,1,1) \nonumber \\
	\textrm{Eigenmode 1} :&~\lambda_1=1-3P,~\ee_1=\frac{1}{6}(2,-1,-1) \nonumber \\
	\textrm{Eigenmode 2} :&~\lambda_2=1-P-2Q,~\ee_2=\frac{1}{2}(0,1,-1).
\label{mc1sol}
\end{align}

Asymptotically ($s\to\infty$), an individual has an equal probability $\pp=\ee_0=(1/3,1/3,1/3)$ to be found at each cell.  Eigenmodes 1 and 2 dictate the transient behavior: According to Eq.~\eref{mc1sol}, in eigenmode 1, cell $1$ has a positive eigenvector component $(\phi_{1,1}>0)$ while $2$ and $3$ have negative components. Thus if $\lambda_1=1-3P>0$, the walker tends to be confined at cell $1$ or in the cluster of cells $2$ and $3$. This is intuitively understandable; a small $P$ means that transition between the two clusters is discouraged, leading to the confinement, and vice versa.  Eigenmode 1 does not give us the details on the transition between $2$ and $3$, however. This finer detail is given by eigenmode 2: In it, cells $2$ and $3$ are in separate clusters (cell $1$ is completely irrelevant with $\ee_{2,1}=0$): if $\lambda_3=1-P-2Q>0$, the walker is confined to either $2$ or $3$ with only a small probability of transitioning to the other, and vice versa. This is presented graphically in  Fig.~\ref{realeig}~(b).

The discussion so far puts us in a position to explain the behaviors of $\cF(\lambda,s)$ we see in Fig.~\ref{Fbehavior}. When $\lambda<0$, the individual has a high propensity to transition to the other cluster than one he is in.  Thus the individual covers a larger area in the same time, and thus the radius of gyration increases faster than in the case of $\lambda>0$.  Therefore, concentration of mode weights $\rho$ on modes with $\lambda\simeq1$ is the manifestation of the nature of relatively local movements of humans -- were the opposite, we should see large $\rho$ on modes that show a faster increase of the radius of gyration.

We present a real-world example of the Voronoi cell clustering in Fig.~\ref{realeig}~(c). It shows a small rectangular portion of the country containing ten Voronoi cells (left)~\footnote{The cells have been slightly deformed to mask the identity of the country}. A segment of the nation's highway runs along the set. We consider two largest eigenmodes, 1 and 2, of an individual who has traversed the cells. $\lambda_1$ and $\lambda_2$ satisfy $1>\lambda_1>\lambda_2$. Similar to the simple example of Fig.~\ref{realeig}~(b), according to eigenmode 1 the entire strand forms a single cluster (center). This agrees well with our intuition regarding a traveler on a highway -- once they enter a highway, it is likely that they stay on it, visiting each cell (along the strand) in succession.  Yet, this is not yet a conclusive proof of it; we cannot exclude the possibility, however unintuitive it may sound, that the user had driven in a haphazard manner among the cells.  The cell clustering according to eigenmode 2 (right) provides an answer: $\lambda_2>0$ means that there could have been only minimal transitions between the two clusters, showing that the traveler did move in a linear fashion on the highway.

\subsection{Complex-valued Eigenmodes}
\begin{figure*}
\resizebox{4.0\figurewidth}{!}{\includegraphics{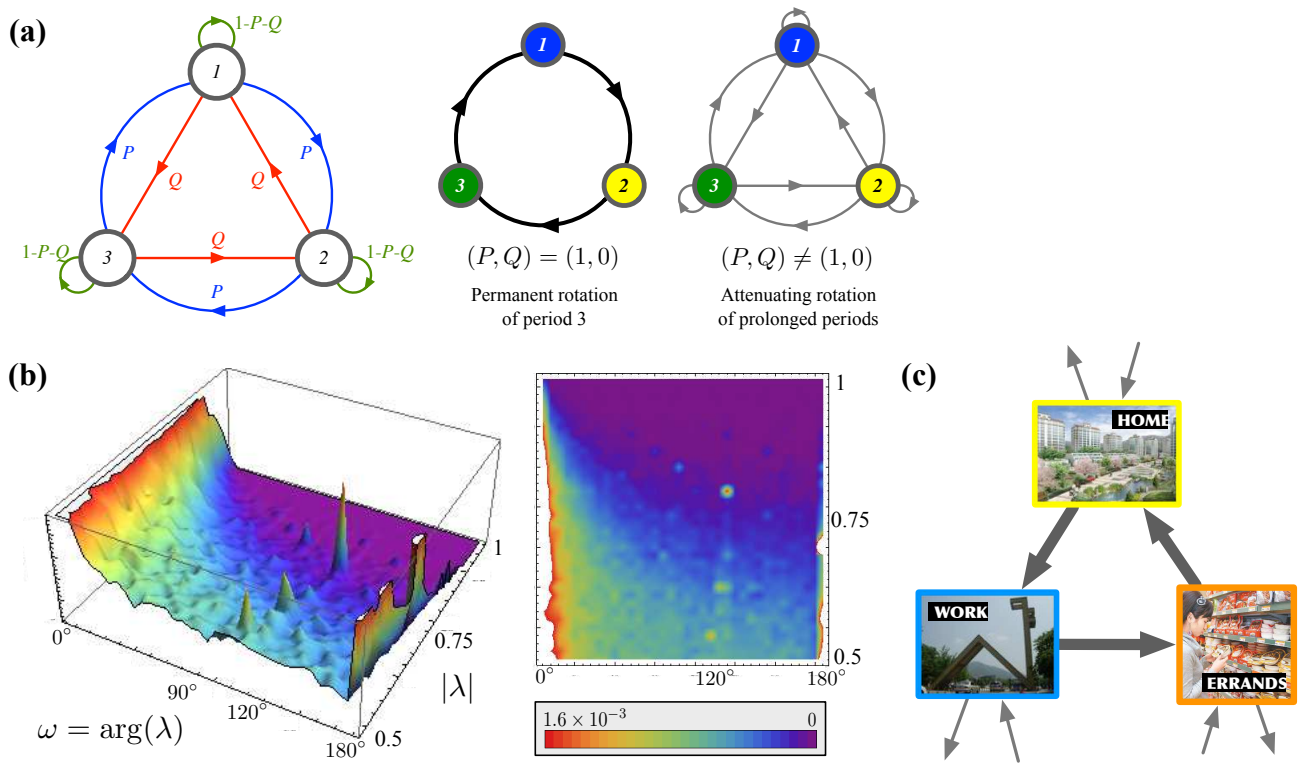}}
\caption{(a) A simple asymmetric Markov process with a natural $120\dgr$ rotational symmetry. According to eigenmodes 1 and 2 (complex conjugates of each other) in Eq.~\eref{mc2sol}, each cell constitutes a distinct cluster, resulting in three separate clusters (blue, yellow, and green in the figure). The dynamic property of the motion depends on $P$ and $Q$. When $(P,Q)=(1,0)$, a permanent (non-attenuating) clockwise rotation of period $3$ occurs. In more general cases of $(P,Q)\ne(1,0)$, a combination of stationary ($1-P-Q$) and dynamic ($P+Q$) tendencies lead to an attenuating rotation of varying prolonged periods. (b) The distribution of $\omega$ for eigenmodes with $|\lambda|>0.5$. In addition to real eigenvalues ($\omega\in\set{0,180\dgr}$), we find  prominent peaks at $\omega\simeq2\pi/3(120\dgr)$, indicating that a rotational motion of (nearly) period three is commonplace. (c) An example trajectory giving $\omega\simeq2\pi/3$ as a prominent eigenmode of motion.}
\label{complexeig}
\end{figure*}

We can readily extend the interpretation of real eigenmodes of the previous section into that of complex-valued eigenmodes; we need to consider that now the angular variables $\set{\alpha_k,\omega_k,\phi_{k,n}}$ in Eq.~\eref{pevolve} can be of any value in $[0,2\pi)$. This means that the Voronoi cells can be grouped into more than two clusters, and that the occupation probabilities of each cell may oscillate with varying periods. We again illustrate this point with a simple example shown in Fig.~\ref{complexeig}~(a), which exhibits a natural $120\dgr$ rotational symmetry. Solving for its transition matrix, we obtain the following eigenmodes (without loss of generality, we assume $P>Q$):
\begin{align}
	\textrm{Eigenmode 0} : &~\lambda_0=1,~\ee_0=\frac{1}{3}(1,1,1) \nonumber \\
	\textrm{Eigenmode 1} : &~\lambda_1=\frac{1}{2}\bigl((2-3(P+Q)+i\sqrt{3}(P-Q)\bigr),\nonumber \\
	&~\ee_1=(\e^{-i2\pi/3},\e^{i2\pi/3},1) \nonumber \\
	\textrm{Eigenmode 2} : &~\lambda_2=\lambda_1^{*},~~\ee_2=\ee_1^{*}
\label{mc2sol}
\end{align}

The asymptotic occupation probability is again $\pp=\ee_0=(1/3,1/3,1/3)$.  More interestingly, we observe that the eigenvector $\ee_1$ and its conjugate $\ee_2$ cluster the cells into three distinct groups, regardless of $P$ and $Q$. The transient dynamics of the traveler determined by $\omega_1$ and $\omega_2$, however, does not always reflect the underlying $120\dgr$ symmetry (i.e., a rotational movement of period three). To see why, let us rewrite $|\lambda_1|$ and $\omega_1=\mathrm{arg}(\lambda_1)$:
\begin{align}
	|\lambda_1|	&= \frac{1}{2}\sqrt{(2-3\delta)^2+3\eta^2} \nonumber \\
	\omega_1	&= \arctan\biggl(\frac{\sqrt{3}\eta}{2-3\delta}\biggr),
\end{align}
where we have defined $\delta\equiv(P+Q)$ and $\eta\equiv (P-Q)$. We can say that $\delta$ represents the dynamism of the traveler since it is the probability to transition to a different cell at each step, while $\eta$ represents the rotational tendency of the traveler since a large $\eta$ would result in a stronger clockwise tendency of movement.
First, assume $P=1$ and $Q=0$ so that $\delta=\eta=1$. This leads to $|\lambda_1|=1$ and $\omega_1=2\pi/3$, and resulting from it is a perpetual (non-attenuating) clockwise rotational motion of period three, perfectly reflecting the symmetry of the diagram.  Any other combination of $P$ and $Q$ leads to a slower, attenuating motion with larger periods, if at all -- from the construction of Fig.~\ref{complexeig}~(a), a non-zero $Q$ or $1-P-Q$ would lead to what would function as ``friction'' against a pure rotation of period three.

Yet, given that a period-three rotation is the simplest periodic motion arising from complex-valued eigenmodes, it appears worth investigating whether such motion can be readily found in real human trajectories.  To check the possibility, we studied the distribution of $\omega=\mathrm{arg}(\lambda)$ for eigenvalues with magnitude $|\lambda|\ge0.5$ (for reasonable persistence), shown in Fig.~\ref{complexeig}~(b).  We find that, although the majority of eigenvalues are real ($\omega=0$ or $\pi$), there exist prominent peaks at $\omega\simeq2\pi/3(=120\dgr)$. In fact, the peaks represent a sizable portion of the population (in our dataset, $24.7\%$ of the individuals)~\footnote{Although it would not be an isolated movement like the one shown in Fig.~\ref{complexeig}~(a), it means that it is sufficiently isolated to exhibit such eigenvalues.}. Although our current data set does not include detail on the exact places of visit or activities undertaken by each individual, we present one scenario that appears to be a reasonable origin of the behavior in Fig.~\ref{complexeig}~(c): It is a schematic of the trajectory of one of the authors (J.P.) during typical weekdays, composed mainly of a Home-to-Work-to-Errands-then-back-to-Home trajectory with only occasional escapes from it. It would be extremely interesting, when data much more detailed than the one analyzed here becomes available, to extract such non-trivial yet pronounced (dominant) eigenmodes of individuals and see whether they match well to the perceived units of motion (e.g., commuting made up of a series of transitions between locales).

\section{Discussion}
\label{sectionD}
In this paper, we have presented in detail how the framework of stochastic processes, widely used in theoretical physics, can be used for analysis of large-scale human motion data. Specifically, utilizing the well-established theory of Markov matrices, we have demonstrated that the observed temporal evolution of the radius of gyration can be understood via eigenmode analysis of individual transition matrices. We have also discussed how the eigenvalues and eigenvectors are related to the microscopic characteristic modes of individual mobility.

We anticipate our approach to grow more relevant as innovations continue in large-scale data acquisition technology: even now, with GPS-enabled mobile phones becoming more available, individuals are able to track their own movements over locales of interest and couple them with digitalized geographical information (the so-called ``geo-tagging'') to construct a detailed space-time history of one's past whereabouts. On a more social scale, furthermore, we can readily imagine the benefits of a better understanding of human mobility patterns: it would allows us to better design infrastructure such as roads, transportation systems, and vital utilities so that social cost is minimized while location-based human activities are optimally supported~\cite{Um:2009}. Also notable is the active research effort in the field of ecology to understand animal movements~\cite{Nathan:2008,Revilla:2008} which have a potential to be helpful in understanding human mobility in urban enviroments as well.

We hope that our work plays a role in highlighting the opportunities for theoretical physicists to make novel and innovative contributions to social and technological problems.

\acknowledgments
We would like to thank A.-L. Barab\'asi and Doochul Kim for helpful suggestions. This work was supported by the Brain Korea 21 Frontier Physics Program, the Future Internet Forum of Korea. Park acknowledges Kyung Hee University grant KHU-20100116 and the National Research Foundation of Korea grant KRF-20100004910. Lee acknowledges the support from the National Research Foundation of Korea grant No. 2009-0063911.


\begin{thebibliography}{99}
\bibitem{Nature:2009} \textit{Community Cleverness Required}, Nature \textbf{455}, 1 (2009)
\bibitem{Brockmann:2006} D. Brockmann, L. Hufnagel and T. Geisel, Nature \textbf{439}, 462 (2006)
\bibitem{Gonzalez:2008} M. Gonzal\'ez, C. Hidalgo and A.-L. Barabasi, Nature \textbf{453}, 779(2008)
\bibitem{Mantegna:1994} R. N. Mantegna and H. E. Stanley, Phys. Rev. Lett. \textbf{73}, 2946(1994)
\bibitem{Klafter:1996} J. Klafter, M. F. Shlesinger and G. Zufomen, Physics Today \textbf{49} 33(1996)
\bibitem{Hughes:1995} B. D. Hughes, \textit{Random Walks in Random Environments Vol.1} (Oxford University Press, Oxford, 1995)
\bibitem{Hughes:1996} B. D. Hughes, \textit{Random Walks in Random Environments Vol.2} (Oxford University Press, Oxford, 1996)
\bibitem{Sinai:1983} Y. G. Sinai, Theo. Prob. Appl. \textbf{27}, 256 (1983)
\bibitem{Durrett:1986} R. Durrett, Commun. Math. Phys. \textbf{104}, 87 (1986)
\bibitem{Kesten:1986} H. Kesten, Physica A \textbf{138}, 299 (1986)
\bibitem{Markov:2006} A. A. Markov, Sci. Context \textbf{19}, 591 (2006)
\bibitem{Shannon:1948} C. E. Shannon, Bell. Sys. Tech. J. \textbf{27}, 379 (1948)
\bibitem{Bremaud:1991} P. Bremaud, \textit{Markov Chains: Gibbs fields, Monte Carlo simulation, and queues} (Springer, New York, 1991)
\bibitem{Krogh:1994} A. Krogh, M. Brown, I. Mian, K. Sjolander and D. Haussler, J. Mol. Biol. \textbf{235}, 1501 (1994)
\bibitem{Candia:2008} J. Candia, M. Gonz\'alez, P. Wang, T. Schoenharl, G. Madey and A.-L. Barabasi, J. Phys. A \textbf{41}, 224015 (2008)
\bibitem{Goldstein:1950} H. Goldstein, \textit{Classical Mechanics} (Addison-Wesley, Reading, 1980)
%\bibitem{Ortuzar} J. D. Ortuzar and L. Willumsen, Modeling Transport, \textit{Wiley} (2001)
%\bibitem{Wilson} A. G. Wilson, \textit{Oper Res Quar} \textbf{21}, 247-265 (1970)
\bibitem{Um:2009} J. Um, S. Son, S. Lee, H. Jeong and B. Kim, Proc. Nat. Acad. Sci. \textbf{106}, 14236 (2009)
\bibitem{Nathan:2008} R. Nathan, W. Getz, E. Revilla, M. Holyoak, R. Kadmon, D. Saltz and P. Smouse, Proc. Nat. Acad. Sci. \textbf{105}, 19052(2008)
\bibitem{Revilla:2008} E. Revilla and T. Wiegand, Proc. Nat. Acad. Sci. \textbf{105}, 19120 (2008)
\end{thebibliography}
\end{document}